\documentclass[12pt,reqno]{amsart}
\usepackage{amsmath}
\usepackage{amsfonts}
\usepackage{amssymb}
\usepackage{amsthm}
\usepackage{newlfont}
\usepackage{epsfig}
\usepackage{amscd}

\theoremstyle{plain}
\newtheorem{Th}{Theorem}
\newtheorem{Cor}[Th]{Corollary}
\newtheorem{Lem}[Th]{Lemma}
\newtheorem{Prop}[Th]{Proposition}
\theoremstyle{definition}
\newtheorem{Def}{Definition}

\theoremstyle{remark}
\newtheorem*{Rem}{Remark}
\newcommand{\PP}{{\mathbb P}}
\newcommand{\GG}{{\mathbb G}}
\newcommand{\RR}{{\mathbb R}}
\newcommand{\EE}{{\mathbb E}}
\newcommand{\ZZ}{{\mathbb Z}}

\newcommand{\bx}{{\boldsymbol x}}

\newcommand{\bN}{{\boldsymbol N}}
\newcommand{\by}{{\boldsymbol y}}
\newcommand{\bu}{{\boldsymbol u}}

\newcommand{\bw}{{\boldsymbol w}}

\newcommand{\vbu}{\vec{\bu}}
\newcommand{\vbN}{\vec{\bN}}
\newcommand{\vbn}{\vec{\boldsymbol n}}
\newcommand{\vbomega}{\vec{\boldsymbol \omega}}
\newcommand{\vb}{\vec{\boldsymbol b}}
\newcommand{\vecbfeta}{\vec{\boldsymbol \eta}}

\begin{document}
\title[The dual {B}ianchi lattice]{The normal dual congruences \\ 
and the dual {B}ianchi lattice}
\author[A. Doliwa]{Adam Doliwa}
\address{Wydzia{\l} Matematyki i Informatyki \\
Uniwersytet Warmi\'nsko--Mazurski w Olsztynie \\
ul. \.Zo{\l}nierska 14A \\ 
10-561 Olsztyn, Poland}
\email{doliwa@matman.uwm.edu.pl} 
\begin{abstract}
The main goal of the paper is to find the discrete analogue
of the Bianchi system in spaces of arbitrary dimension
together with its geometric interpretation.
We show that the proper geometric framework of such generalization 
is the language of dual quadrilateral lattices and of dual congruences. 
After introducing the notion of
the dual Koenigs lattice in a projective space of arbitrary dimension
we define the discrete dual congruences and we present, as an important 
example, the normal dual discrete congruences. Finally, we introduce
the dual Bianchi lattice as a dual Koenigs lattice allowing for
a conjugate normal dual congruence, and we find 
its characterization in terms of a
system of integrable difference equations. 
\end{abstract}
\maketitle
\section{Introduction}
In paper \cite{Bianchi-BN} Bianchi studied
properties of the Moutard equation \cite{Moutard}
\begin{equation*}
\frac{\partial^2 \vbN}{\partial u \partial v}=f\vbN,
\end{equation*}
subjected to the quadratic reduction
\begin{equation*}
\frac{\partial^2 \vbN}{\partial u \partial v}=f\vbN, \qquad
\vbN\cdot\vbN = U + V;
\end{equation*}
here $f$ is a given function of two variables $u,v$, while $U$ and $V$ are
functions of single variables $u$ and $v$, correspondingly. The unknown
function $\vbN$ takes values in the space $\RR^M$ equipped with the scalar
product "$\cdot$". 

In the basic case of $M=3$ the above system is known as
the Bianchi system and is related with the Ernst reduction of Einstein's
equation (see the relevant literature in \cite{Nieszporski-mE}, where such
correpondence was also generalized to the case of $M>3$).
In the classical differential geometry the Bianchi system for $M=3$
appears in the contexts of: \\
(i) conjugate nets permanent in deformation \cite{Eisenhart-TCS};\\
(ii) conjugate nets with equal tangential invariants allowing for
harmonic normal congruence \cite{Eisenhart-TS};\\
(iii) asymptotic nets allowing for the Weingarten
transformations preserving the Gauss
curvature \cite{Bianchi}.\\
In the interpretations (i) and (ii), which in fact describe the same
geometric object from different points of view, the variables $u,v$ are 
conjugate parameters. In the case (iii) the variables $u,v$ are the
asymptotic coordinates on a hyperbolic surface.

The Bianchi system is, due to its physical importance, one of the most
interesting soliton systems having the distinguished geometric meaning.
During recent years many of the geometrically meaningful integrable systems
have found their integrable discrete analogues (see, for example, 
\cite{DS-AL,MQL,BobenkoPinkall-DSIS,KoSchief2,TQL,DS-sym,NieszporskiA} or a 
recent review \cite{Dol-IMDG}). 

The discrete integrable analogue of the Bianchi system for $M=3$ has been
obtained in \cite{NDS,DNS-I} as a reduction of the discrete Moutard
equation \cite{NiSchief} which describes asymptotic lattices
\cite{NieszporskiA,W-cong}, thus via discrete version of the interpretation
(iii). It turns out that the same equation was obtained in the
earlier work \cite{Schief-C}, for arbitrary $M$, in a different geometric context
of isothermic lattices \cite{BP2}.

An alternative approach, based on the interpretation (ii), 
to the integrable discretization of the Bianchi system
(again for $M=3$) was presented in \cite{DNS-gdBs}. Suprisingly, one obtains
in this way not the system derived in \cite{NDS,DNS-I}, but an integrable
reduction of the Koenigs lattice equation \cite{Nieszporski-Lapl,Dol-Koe};
there exists, however, a simple relation between both discrete 
versions of the Bianchi system. We remark that also the interpretation (i) 
has been recently discretized~\cite{Schief-Arran}. 
  
The algebraic transition to $M>3$ 
of the discrete Bianchi system obtained in \cite{DNS-gdBs} does not
need special efforts. The main goal of the paper is to find the proper geometric
framework for such a generalization. We define therefore, in addition to 
recently introduced notion of the dual quadrilateral lattice \cite{DS-sym}, the
notion of the discrete dual congruence, thus making the theory of quadrilateral
lattices \cite{MQL} and their transformations \cite{TQL} perfectly dual (in the
sense of the standard duality in projective spaces). 

The layout of the paper is as follows. We first recall the definition 
of the dual quadrilateral lattice and we introduce 
the dual Koenigs lattice in a projective space of arbitrary dimension. 
In Section \ref{sec:dual_congruences} we define the discrete 
dual congruences and we study their basic properties. The next
Section~\ref{sec:normal_dual_congruences}  
is devoted to an important class of such congruences, the normal dual 
discrete congruences. Then, in Section \ref{sec:dual_Bianchi} we introduce
the dual Bianchi lattice as a dual Koenigs lattice allowing for
a conjugate normal dual congruence. Finally, we characterize the dual
Bianchi lattices in terms of solutions of the discrete Bianchi system and we
provide the Darboux-type transformation for the system.

\section{The dual Koenigs lattice}
The dual quadrilateral
lattices, introduced in \cite{DS-sym}, are lattices of planar 
quadrialterals in the dual space $(\PP^M)^*$. \begin{Def}
A map $x^*:\ZZ^2\to (\PP^M)^*$, $M> 2$, from two dimensional integer lattice
into the space of hyperplanes in $M$-dimensional projective space $\PP^M$
is called a {\em dual quadrilateral lattice} if the intersection of
any four neighbouring
hyperplanes $x^*$, $x^*_{(1)}$, $x^*_{(2)}$ and $x^*_{(12)}$ is 
a projective subspace of co-dimension three.
\end{Def}
By subscripts in brackets we
denote shifts in the corresponding discrete variables, for example
$f_{(\pm1)}(m_1,m_2)=f(m_1\pm 1,m_2)$. 
\begin{Rem}
We will consider here only non-degenerate dual quadrilateral lattices for 
which no three of the four neighbouring hyperplanes have intersection of
co-dimension two. 
\end{Rem}
\begin{Rem}
The case of $M=3$ is special, because any quadrilateral lattice 
$x:\ZZ^2\to\PP^3$ gives rise to a dual quadrilateral lattice 
$x^*:\ZZ^2\to (\PP^3)^*$, where $x^*$ is the plane passing through 
the points $x$, $x_{(1)}$ and $x_{(2)}$. 
\end{Rem}
The homogeneous coordinates $\bx^*:\ZZ^2\to\RR^{M+1}\setminus \{0\}$ 
of the dual quadrilateral lattices are subjected to the 
discrete Laplace equation
\begin{equation} \label{eq:Laplace-dis-hom-2-dual}
\bx^*_{(12)} = A^*_{(1)}\bx^*_{(1)} + B^*_{(2)}\bx^*_{(2)} + C^*\bx^*.
\end{equation}
Denote by
$L^*_i$ the subspace of co-dimension two (a dual line) being the
intersection of $x^*$ and $x^*_{(i)}$, $i=1,2$. The hyperplanes containing
$L^*_i$ (such a pencil of hyperplanes is often identified with $L^*_i$) have
coordinates $\lambda \bx^* + \mu \bx^*_{(i)}$. The hyperplanes $x^*_{-1}$
with homogeneous coordinates 
\begin{equation} \label{eq:x^*_{-1}}
\bx^*_{-1} = \bx^*_{(1)}-B^*\bx^* = A^*_{(1-2)}\bx^*_{(1-2)} + C^*_{(-2)}
\bx^*_{(-2)},
\end{equation}
are contained in the corresponding pencils $L^*_1$ and
$L^*_{1(-2)}$.
Similarly, the hyperplanes $x^*_{1}$
with homogeneous coordinates 
\begin{equation} \label{eq:x^*_{1}}
\bx^*_{1} = \bx^*_{(2)}-A^*\bx^* = B^*_{(-1 2)}\bx^*_{(-1 2)} + C^*_{(-1)}
\bx^*_{(-1)},
\end{equation}
are contained in the pencils $L^*_2$ and
$L^*_{2(-1)}$.
\begin{Rem}
Above we defined the Laplace transforms~\cite{DCN} of dual quadrilateral 
lattices.
\end{Rem}

A distinguished example of the integrable dual lattice is the dual Koenigs 
lattice (for definition of the Koenigs lattice
see \cite{Dol-Koe}), introduced in an equivalent form
in the special case $M=3$ in \cite{DNS-gdBs}.
\begin{Def}
A dual quadrilateral lattice $x^*:\ZZ^2\to (\PP^M)^*$ is called a {\em dual 
Koenigs lattice} if the three pencils: (i) containing the hyperplanes
$x^*_1$ and $x^*_{-1(22)}$, (ii) containing the hyperplanes
$x^*_{-1}$ and $x^*_{1(11)}$, (iii) containing the hyperplanes
$x^*$ and $x^*_{(12)}$, have a hyperplane in common.
\end{Def}
\begin{Rem}
For generic dual quadrilateral lattice these three pencils contain the
projective subspace $x^*\cap x^*_{(1)} \cap x^*_{(2)}$ of co-dimension 
three.
\end{Rem}
To characterize algebraically the dual Koenigs lattices we present the
corresponding reduction of the general discrete Laplace equation for dual
quadrilateral lattices. 
\begin{Th}
A dual quadrilateral lattice $x^*:\ZZ^2\to (\PP^M)^*$ is a dual Koenigs
lattice if and only if its Laplace
equation can be gauged into the form
\begin{equation} \label{eq:Koenigs-dis-dual}
\bx^*_{(12)} + \bx^* = F^*_{(1)}\bx^*_{(1)} + F^*_{(2)}\bx^*_{(2)}.
\end{equation}
\end{Th}
\begin{proof}
We will follow the reasoning of \cite{DNS-gdBs}. 
There exists a common hyperplane of the three pencils if and only if the 
linear system 
\begin{equation} \label{eq:lin-eqs-F^*}
\lambda \bx^*_1 + \mu \bx^*_{-1(22)} = \sigma \bx^*_{-1} + 
\rho\bx^*_{1(11)}
= \chi\bx^* + \nu \bx^*_{(12)}
\end{equation}
has a non-trivial solution $\lambda,\mu ,\sigma,\rho,\chi,\nu$. Using 
equations
\eqref{eq:x^*_{-1}}-\eqref{eq:x^*_{1}} one can rewrite the linear equations 
\eqref{eq:lin-eqs-F^*} in the form
\begin{equation*} \begin{split}
\lambda(\bx^*_{(2)}  - & A^*\bx^*) + 
\mu(A^*_{(12)}\bx^*_{(12)} + C^*_{(2)}\bx^*_{(2)}) = \\
& \sigma(\bx^*_{(1)} - B^* \bx^*) +
\rho(B^*_{(12)}\bx^*_{(12)} + C^*_{(1)}\bx^*_{(1)}) = 
\chi\bx^* + \nu \bx^*_{(12)}.
\end{split} \end{equation*}
The with the help of the Laplace equation 
\eqref{eq:Laplace-dis-hom-2-dual} one concludes that such a solution exists if
and only if the coefficients of the Laplace equation are restricted by the
following equation 
\begin{equation*}
A^*C^*_{(2)}B^*_{(12)} = B^* C^*_{(1)} A^*_{(12)}.
\end{equation*}
This restriction on the coefficients of the Laplace
equation~\eqref{eq:Laplace-dis-hom-2-dual} implies~\cite{Dol-Koe} 
existence of the gauge function
$\rho$ defined by
\begin{equation*}
\rho_{(12)} = -C^*\rho, \qquad \rho_{(1)} A^* = \rho_{(2)}B^*.
\end{equation*}
After the gauge transformation $\bx^*\mapsto\bx^* /\rho$, the new 
coordinates of the dual lattice satisfy the Laplace equation of the
form~\eqref{eq:Koenigs-dis-dual} with the potential
\begin{equation*}
F^*=\frac{A^*\rho}{\rho_{(2)}} = \frac{B^*\rho}{\rho_{(1)}}.
\end{equation*} 
\end{proof}
In \cite{Dol-Koe} there was also found the reduction of the fundamental 
transformation acting within the class of lattices whose homogeneous coordinates
satisfy 
equation \eqref{eq:Koenigs-dis-dual}. The geometric content of the following 
result will be presented in the next Section.
\begin{Prop} \label{th:transf-dKd}
Given the dual Koenigs lattice $x^*:\ZZ^2\to (\PP^M)^*$
with homogeneous coordinates $\bx^*$ satisfying equation 
\eqref{eq:Koenigs-dis-dual} 
and given a scalar solution $\theta$ of its adjoint equation (the Moutard 
equation) 
\begin{equation} \label{eq:Moutard-dis}
\theta_{(12)}+\theta = F^*(\theta_{(1)}+\theta_{(2)}),
\end{equation}
define the functions
\begin{equation} \label{eq:phi}
\phi = \theta_{(1)} + \theta_{(2)}, \qquad \phi^\prime = 
\frac{1}{\theta_{(1)}} + \frac{1}{\theta_{(2)}}.
\end{equation}
Then the solution $\bx^{*\prime}$ of the linear system
\begin{eqnarray} \label{eq:x-x'-1}
\Delta_1\left( \frac{\bx^{*\prime}}{\phi^\prime} \right) & = &\;
 (\theta \theta_{(2)})_{(1)}\Delta_1\left( \frac{\bx^*}{\phi} \right), \\
\label{eq:x-x'-2} 
\Delta_2\left( \frac{\bx^{*\prime}}{\phi^\prime} \right) & = &
- (\theta \theta_{(1)})_{(2)}\Delta_2\left( \frac{\bx^*}{\phi} \right),
\end{eqnarray}
satisfies equation \eqref{eq:Koenigs-dis-dual} with
\begin{equation} \label{eq:F'}
F^{*\prime}=F^*\frac{\theta_{(1)}\theta_{(2)}}{\theta\theta_{(12)}},
\end{equation}
and defines homogeneous coordinates of the new dual Koenigs lattice.
\end{Prop}
\begin{Rem}
The Koenigs lattice equation \eqref{eq:Koenigs-dis-dual} was introduced  in
\cite{Nieszporski-Lapl} in the context of the 
Laplace sequences of quadrilateral lattices. 
\end{Rem}

\section{Dual congruences} \label{sec:dual_congruences}
\begin{figure}
\begin{center}
\leavevmode\epsfysize=5cm\epsffile{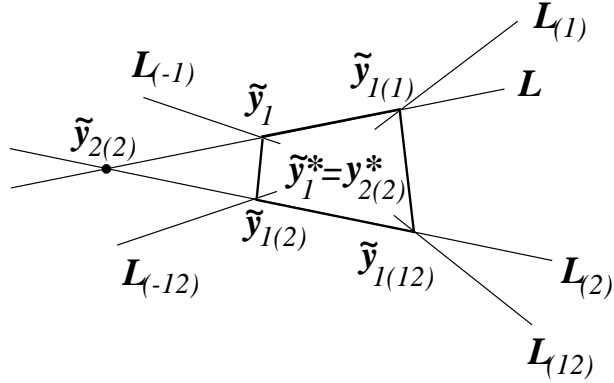}
\end{center}
\caption{The focal planes of a congruence in $\PP^3$}
\label{fig:focal-planes}
\end{figure}

In this Section we introduce the dual version of
the second basic object of the discrete integrable 
geometry, the notion of a discrete dual congruence.
In what follows by $\GG(K+1,M+1)$ we denote the set of $(K+1)$-dimensional
linear subspaces of the
$(M+1)$-dimensional vector space $\RR^{M+1}$, thus the space of $K$-dimensional
projective subspaces of the $M$-dimensional projective space $\PP^M$.
\begin{Def}
A map $L^*:\ZZ^2\to\GG(M-1,M+1)$, $ M> 2$, is called
a {\em dual congruence} if any two
neighbouring subspaces are contained in a subspace of co-dimension one.
The lattices of 
common hyperplanes $y^*_i:\ZZ^2\to (\PP^M)^*$, $i=1,2$, of 
$L^*$ and $L^*_{(-i)}$ are called {\em focal hyperplane lattices} (or {\em dual
focal lattices})
of the dual congruence.  
\end{Def}
\begin{Rem}
We will consider here only non-degenerate dual congruences for 
which any subspace of the congruence cannot have common hyperplanes 
with other subspaces different from its nearests neighbours.
\end{Rem}
\begin{Cor}
The Grassmann space $\GG(M-1,M+1)$ is the natural dual of the space 
$\GG(2,M+1)$ of lines in $\PP^M$. Moreover, the dual version of two
intersecting lines are two projective subspaces of co-dimension two
contained in a hyperplane. Therefore, all known results of the theory of 
discrete congruences~\cite{TQL} have the natural dual counterpart.
\end{Cor}
In particular, given dual quadrilateral lattice $x^*:\ZZ^2\to (\PP^M)^*$ then
the maps $L_i^*:\ZZ^2\to\GG(M-1,M+1)$, $i=1,2$, defined by 
$L^*_i= x^*\cap x^*_{(i)}$ and considered in the previous
Section, are dual congruences. 
They are the dual versions of the tangent congruences 
$L_i:\ZZ^2\to\GG(2,M+1)$ of a quadrilateral
lattice $x:\ZZ^2\to \PP^M$, which are given by lines $L_i$ passing through
$x$ and $x_{(i)}$.
\begin{Rem}
In the case $M=3$ the notion of line is self-dual and any congruence is
simultaneously a dual congruence as well. However, if 
$x:\ZZ^2\to \PP^3$ is a quadrilateral lattice and 
$x^*:\ZZ^2\to (\PP^3)^*$ is the dual quadrilateral lattice of its tangent
planes, then $L^*_1=L_{2(1)}$ and $L^*_2=L_{1(2)}$.
\end{Rem} 
Analogously, we have in the case $M=3$ the following
double interpretation of "focal planes" of the congruence, which had to be
taken into account in \cite{DNS-gdBs} (see Figure~\ref{fig:focal-planes}).
\begin{Cor} \label{cor:focal-planes}
Let $L=L^*:\ZZ^2\to\GG(2,4)$ be a congruence in $\PP^3$ and  
let $y_i^*:\ZZ^2\to (\PP^3)^*$, $i=1,2$, denote  its dual focal lattices.
If by $\tilde{y}^*_i:\ZZ^2\to (\PP^3)^*$ denote the dual lattices
of tangent planes of the focal (point) lattices 
$\tilde{y}_i:\ZZ^2\to \PP^3$, $i=1,2$, of the congruence, then
$\tilde{y}^*_1=y^*_{2(2)}$ and $\tilde{y}^*_2=y^*_{1(1)}$.
\end{Cor}
\begin{Prop}
If $L^*:\ZZ^2\to\GG(M-1,M+1)$ is a dual congruence then its
focal hyperplane lattices $y^*_i:\ZZ^2\to (\PP^M)^*$, $i=1,2$, are 
quadrilateral
hyperplane lattices. 
\end{Prop}
\begin{proof}
We will demonstrate the Proposition for the first focal dual lattice.
Denote by $\by^*_1:\ZZ^2\to\RR^{M+1}\setminus \{0\}$, its homogeneous
representants. 
Because both hyperplanes $y^*_{1}$ and $y^*_{1(1)}$ contain the subspace 
$L^*$
then any other hyperplane of this pencil has homogeneous coordinates of 
the form
\begin{equation*}
\alpha \by^*_{1} + \beta \by^*_{1(1)}.
\end{equation*}
Similarly, the homogeneous coordinates of any hyperplane passing through 
$L^*_{(2)}$  are linear combination of $\by^*_{1(2)}$ and $\by^*_{1(12)}$
\begin{equation*}
\gamma \by^*_{1(2)} + \delta \by^*_{1(12)}.
\end{equation*}
Because the hyperplane $y^*_{2(2)}$ is contained in both pencils then for 
some
non-zero $\lambda$ and not all vanishing $\alpha,\beta,\gamma,\delta$
\begin{equation*}
\lambda(\alpha \by^*_{1} + \beta \by^*_{1(1)}) = 
\gamma \by^*_{1(2)} + \delta \by^*_{1(12)}.
\end{equation*} 
\end{proof}
\begin{Cor}
Notice that $\delta=0$ would imply that $L^*$ is contained in $y^*_{1(2)}$, 
which violates the assumption of genericity of the dual congruence.
\end{Cor}
\begin{Rem}
The notion of the dual congruences can be naturally extended to $N$-parameter 
families of dual
lines. The proof that the focal hyperplane lattices are quadrilateral needs
then however additional genericity assumptions (see \cite{TQL}
where the analogous problem for congruences of lines was treated in 
detail). 
\end{Rem}

The dual congruences play the same role in the theory of 
transformations of
quadrilateral dual lattices as the congruences of lines in the theory of 
transformations of quadrilateral lattices.
\begin{Def}
A hyperplane quadrilateral lattice $x^*:\ZZ^2\to (\PP^M)^*$ and a dual
congruence $L^*:\ZZ^2\to\GG(M-1,M+1)$ are said to be {\em conjugate} to each
other if the subspaces $L^*$ of the dual congruence are 
contained in the corresponding hyperplanes $x^*$ 
of the lattice.
\end{Def} 
\begin{Rem}
For $M=3$ when the notion of a congruence is self-dual, 
if $x:\ZZ^2\to  \PP^3$ is a quadrilateral lattice and 
$x^*:\ZZ^2\to (\PP^3)^*$ is the dual quadrilateral lattice of its tangent
planes, then a (dual) congruence $L^*=L:\ZZ^2\to\GG(2,4)$ conjugate to $x^*$ is
called {\em harmonic} to $x$.
\end{Rem}
\begin{Def}
Two quadrilateral dual lattices are related by the {\em dual
fundamental transformation} if
they are conjugated to the same dual congruence.
\end{Def}
The algebraic results of the theory of transformations of the
quadrilateral lattices can be interpreted in the dual space. In particular, 
we present here the dual geometric content of the formulas, originally obtained
for the Koenigs lattices in \cite{Dol-Koe}. The restriction of the dual
fundamental transformation to the class of the dual Koenigs lattices is
algebraically formulated in Theorem~\ref{th:transf-dKd}.     
\begin{Cor} \label{cor:focal-lat-Koe}
Let $\bx^*:\ZZ^2\to\RR^{M+1}\setminus \{ 0\}$ and 
$\bx^{*\prime}:\ZZ^2\to\RR^{M+1}\setminus \{ 0\}$ be the homogeneous 
representants
of two dual Koenigs lattices related by the dual Koenigs transformation
described in Theorem~\ref{th:transf-dKd}. 
The homogeneous representants of the
dual focal lattices of the dual congruence of that transformation 
read
\begin{equation} \label{eq:focal-lat-Koe-d}
\by^*_1 = \theta\theta_{(2)}\frac{\bx^*}{\phi} - 
\frac{\bx^{*\prime}}{\phi^\prime}, \qquad
\by^*_2 = -\theta\theta_{(1)}\frac{\bx^*}{\phi} -
\frac{\bx^{*\prime}}{\phi^\prime}.
\end{equation}
Moreover, they can be found from equations  
\begin{align}
\by^*_1 - \by^*_2&=\theta\bx^*, \nonumber\\
\Delta_1\by^*_2 &= (F^*\bx^*-\bx^*_{(1)})\theta_{(1)}, 
\label{eq:y1y2}\\
\Delta_2 \by^*_1 &= 
(\bx^*_{(2)}-F^*\bx^*)\theta_{(2)} \nonumber.
\end{align}
\end{Cor}
Given a dual congruence $L^*:\ZZ^2\to\GG(M-1,M+1)$ 
with dual focal lattices $y^*_i:\ZZ^2\to (\PP^M)^*$, $i=1,2$. In the pencil
of hyperplanes containing $L^*$ define the projective involution 
${\mathfrak i}_{L^*}$ with two pairs of homologous hyperplanes 
$y^*_{i(i)}= {\mathfrak i}_{L^*}(y^*_{i})$, $i=1,2$. The dual Koenigs lattices 
can be geometrically selected from generic
quadrilateral lattices by the following property, being the dual version 
of Corollary~15 of \cite{Dol-Koe}.
\begin{Prop} \label{prop:Koenigs*involution}
A dual quadrilateral lattice $x^*:\ZZ^2\to (\PP^M)^*$ is a dual Koenigs
lattice if and only if, for arbitrary non-degenerate dual
congruence $L^*:\ZZ^2\to\GG(M-1,M+1)$ conjugate to the lattice,
the dual lattice $x^{*\prime}:\ZZ^2\to (\PP^M)^*$ defined by
$x^{*\prime}={\mathfrak i}_{L^*}(x)$ is quadrilateral as well. 
\end{Prop} 
\begin{Cor} \label{cor:y*+-}
In notation of Theorem~\ref{th:transf-dKd} and 
Corollary~\ref{cor:focal-lat-Koe} the homogeneous coordinates
of the fixed hyperplanes $y^*_\pm$ of the involutions 
${\mathfrak i}_{L^*}$ read
\begin{equation} \label{eq:fixed-x-xK}
\by^*_\pm=\pm\sqrt{\theta\theta_{(12)}}\bx^* + 
\sqrt{\theta_{(1)}\theta_{(2)}}\bx^{*\prime}. 
\end{equation}
\end{Cor}

\section{The normal dual congruences} \label{sec:normal_dual_congruences} 
\begin{figure}
\begin{center}
\leavevmode\epsfysize=7cm\epsffile{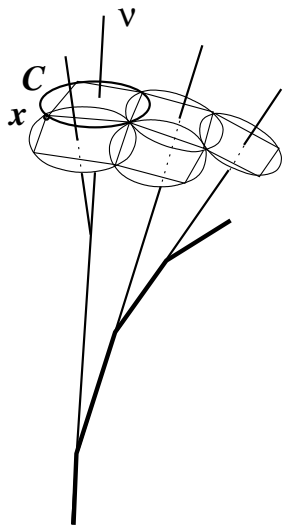}
\end{center}
\caption{The normal congruence of a circular lattice in $\EE^3$}
\label{fig:dis-norm-cong}
\end{figure}

In this Section we define the normal dual congruences. The normal 
congruences in
the (self-dual) case $M=3$ were defined in \cite{DNS-gdBs} following the 
example (see \cite{Dol-Rib}) 
of normal congruences of circular lattices in three 
dimensional Euclidean space $\EE^3$. 
The circular lattice is a quadrilateral
lattice whose quadrilaterals can be inscribed in circles, and its properties
has been studied in detail 
in~\cite{Bobenko-O,CDS,KoSchief2,DMS,LiuManas,q-red,AKV,DMM-KP}.
In \cite{Dol-Rib} it was shown that the normals $\nu$ to
circles of the circular lattice $x:\ZZ^2 \to \EE^3$
passing through the centers of circles (see
Figure~\ref{fig:dis-norm-cong}) form a congruence.
In \cite{DNS-gdBs} such congruences were charcterized 
without any reference to a circular lattice, which was then used to 
define normal congruences in $\EE^3$ as follws
(by $\GG_A(K+1,M+1)$ we denote the set of $K$-dimensional (affine) 
subspaces of the $M$-dimensional Euclidean space $\EE^M$).  
\begin{Def} \label{def:norm-cong-3}
A congruence $\nu:\ZZ^2\to\GG_A(2,4)$ with focal lattices 
$\tilde{y}_i:\ZZ^2\to\EE^3$, $i=1,2$, is a {\em normal congruence} 
if the pair of orthogonal planes bisecting the angles between
the planes $\tilde{y}_{1(-2)}^*$ and $\tilde{y}_1^*$ coincides with 
those bisecting $\tilde{y}_{2(-1)}^*$ and $\tilde{y}_2^*$ .
\end{Def}
Taking into account self-duality of lines in $\PP^3$ and
Corollary~\ref{cor:focal-planes}, one can extend the above definition
to $M>3$.
\begin{Def}
A dual congruence $\nu^*:\ZZ^2\to\GG_A(M-1,M+1)$ with dual focal lattices 
$y^*_i:\ZZ^2\to\EE^M$, $i=1,2$, is a {\em normal dual congruence} 
if the pair of orthogonal hyperplanes bisecting the angles between
the hyperplanesplanes $y^*_{2}$ and $y^*_{2(2)}$ coincides with 
those bisecting $y^*_1$ and $y^*_{1(1)}$ .
\end{Def}
\begin{Cor}
Given a circular lattice $x:\ZZ^2\to\EE^M$, then the unique subspaces $\nu^*$ of
co-dimension two, orthogonal to the planes of circles of the lattice
and containing the centers of the circles, form a normal dual congruence. 
\end{Cor}
Denote by $y_{\pm}^*$ the pair of hyperplanes containing the subspace 
$\nu^*$ 
of a normal dual congruence $\nu^*:\ZZ^2\to\GG_A(M-1,M+1)$ and
bisecting the pair of hyperplanes $y^*_{2}$ and $y^*_{2(2)}$ 
(or $y^*_1$ and $y^*_{1(1)}$). Denote by ${\mathfrak r}_{\nu^*}$ the
unique reflection in the pencil of hyperplanes containing $\nu^*$ such that 
$y^*_{2}$ and $y^*_{2(2)}$ are homologous, i.e. 
${\mathfrak r}_{\nu^*}(y^*_{2})=y^*_{2(2)}$. Then the hyperplanes
$y_{\pm}^*$ are the fixed hyperplanes of the reflection and also
$y^*_{1}$ is homologous to $y^*_{1(1)}$. 

Recall that, within all projective involutions in a pencil of hyperplanes in
$\EE^M$, the reflections are characterized by the property that the 
fixed hyperplanes of the involution are orthogonal. This gives the 
following characterization of the normal dual congruences.
\begin{Prop} \label{prop:circ-cong-ref}
Given a dual congruence $L^*:\ZZ^2\to\GG_A(M-1,M+1)$ with dual focal 
lattices $y^*_i:\ZZ^2\to\EE^M$, $i=1,2$. In the pencils
of hyperplanes containing the dual lines $L^*$ of the dual congruence
consider the projective involutions ${\mathfrak i}_{L^*}$, with two 
pairs of homologous hyperplanes $y^*_{i(i)}= {\mathfrak i}_{L^*}(y^*_{i})$.
The congruence is normal if and only if the fixed hyperplanes of the 
involutions are orthogonal. 
\end{Prop}

\section{The dual quadrilateral Bianchi lattice} \label{sec:dual_Bianchi}
The following definition is a generalization, to the dimension $M\geq3$ of the
ambient space, of the definition of quadrilatertal Bianchi lattice given
in \cite{DNS-gdBs} for $M=3$. 
\begin{Def}
The {\em dual quadrilateral Bianchi lattice} is a dual Koenigs lattice in 
$\EE^M$ allowing for a conjugate normal dual congruence.
\end{Def}
Given a hyperplane $w^*$ in $\EE^M$ then its homogeneous coordinates
$\bw^*=(\vbomega,\omega^{M+1})$ are given up to a multiplication by a
non-zero factor. Then $\vbomega$ is a vector normal (not neccessarily unit)
to the hyperplane
and the point $u\in\EE^M$ with coordinates $\vbu\in\RR^M$ 
belongs to the plane $w$
if $\vbomega\cdot\vbu+\omega^{M+1}=0$. 
\begin{Cor} \label{cor:normal-Bianchi-gauge}
A dual quadrilateral lattice $x^*:\ZZ^2\to\GG_A(M,M+1)$ is a dual 
Koenigs lattice if and only if the field of normal
vectors $\vbn:\ZZ^2\to\EE^M$ to hyperplanes of the lattice can be chosen 
in such a way that it satisies the discrete Koenigs equation
\eqref{eq:Koenigs-dis-dual}.
\end{Cor}
Let us formulate the central result of this Section. The idea of its proof
follows the corresponding geometric reasoning of \cite{DNS-gdBs}.  
\begin{Th} \label{prop:dBianchi-alg}
The dual lattice $x^*:\ZZ^2\to\GG_A(M,M+1)$ 
is a Bianchi lattice if and only if
its normal vector  $\vbn:\ZZ^2\to\EE^M$
can be chosen in such a way that it satisfies the 
discrete Koenigs equation \eqref{eq:Koenigs-dis-dual}
with the potential $F^*:\ZZ^2\to\RR$ subjected to the condition
\begin{equation} \label{eq:dBianchi-ddKoenigs}
\Delta_1\Delta_2 \left( \vbn\cdot\vbn \; F^*\right) = 0.
\end{equation}
\end{Th}
\begin{proof}
$\Rightarrow$ Consider the dual Bianchi lattice $x^*$ and
choose its homogeneous coordinates $\bx^*=(\vbn,x^{*(M+1)})$ satisfying
the discrete Koenigs equation \eqref{eq:Koenigs-dis-dual}. 
Let $L^*$ be a normal dual congruence conjugate to $x^*$, let $y_i^*$,
$i=1,2$, denote the focal dual lattices of the congruence, then by 
Theorem~\ref{th:transf-dKd}, Corollary~\ref{cor:focal-lat-Koe} and
Proposition~\ref{prop:Koenigs*involution}
there exists a solution $\theta$ of the discrete Moutard equation
\eqref{eq:Moutard-dis} and the homogeneous coordinates $(\vbn_i,y_i^{*(M+1)})$ of
the focal lattices can be chosen such that they satisfy the system
\eqref{eq:y1y2}. 

By ${\mathfrak i}_{L^*}$ denote the involutions in pencils of the dual
congruence, defined by conditions $y^*_{i(i)}={\mathfrak i}_{L^*}(y_i)$, 
$i=1,2$. Then the corresponding
dual Koenigs lattice $x^{*\prime}={\mathfrak i}_{L^*}(x^*)$, 
has the homogeneous coordinates $(\vbn^{\prime},x^{*\prime (M+1)})$ 
satisfying the discrete Koenigs equation \eqref{eq:Koenigs-dis-dual} with 
the potential $F^{*\prime}$ given by equation \eqref{eq:F'}. Notice that
because $x^{*\prime}$ is conjugate to the normal dual congruence $L^*$,
then $x^{*\prime}$ is a Bianchi lattice as well.
 
According to Corollary \ref{cor:y*+-} the fixed hyperplanes $y^*_\pm$ of the
involution ${\mathfrak i}_{L^*}$ have the normal vectors given by
\begin{equation} \label{eq:fixed-x-xK-dual-af}
\vbn_\pm =\pm\sqrt{\theta\theta_{(12)}}\vbn + 
\sqrt{\theta_{(1)}\theta_{(2)}}\vbn^\prime . 
\end{equation}
Because the dual $L^*$ congruence is normal then  
Proposition~\ref{prop:circ-cong-ref} implies that 
\begin{equation*}
\vbn_{+}\cdot\vbn_{-}=0,
\end{equation*}
which, due to equations \eqref{eq:fixed-x-xK-dual-af} and  \eqref{eq:F'},
gives
\begin{equation}\label{eq:constraint-dis-1}
\vbn\cdot \vbn \: F^* = \vbn^{\prime} \cdot \vbn^{\prime} \: F^{*\prime} .
\end{equation}

Equations \eqref{eq:phi}, \eqref{eq:focal-lat-Koe-d} and 
\eqref{eq:constraint-dis-1} imply that
\begin{equation} \label{eq:squares-n_1n_2n}
\frac{\vbn_{1}\cdot \vbn_{1}}{\theta\theta_{(2)}} +
\frac{\vbn_{2}\cdot \vbn_{2}}{\theta\theta_{(1)}} = \vbn\cdot\vbn \: F^*.
\end{equation}
Moreover, the same equations supplemented by the transformation formulas 
\eqref{eq:x-x'-1}-\eqref{eq:x-x'-2} give conditions
\begin{equation*}
\Delta_1 \left( \frac{\vbn_{1}\cdot 
\vbn_{1} }{\theta\theta_{(2)}} \right) =0, \qquad 
\Delta_2 \left( \frac{\vbn_{2}\cdot 
\vbn_{2} }{\theta\theta_{(1)}} \right) =0, 
\end{equation*}
which together with equation \eqref{eq:squares-n_1n_2n}
lead to the constraint \eqref{eq:dBianchi-ddKoenigs}.

$\Leftarrow$ By Corollary~\ref{cor:normal-Bianchi-gauge} the dual lattice
$x^*$ is a dual Koenigs lattice. Denote by $\bx^*=(\vbn,x^{*(M+1)})$ its
homogeneous coordinates satisfying the discrete Koenigs 
equation~\eqref{eq:Koenigs-dis-dual}. 
The constraint \eqref{eq:dBianchi-ddKoenigs} implies that 
there exist functions
$U_1$ and $U_2$ of the single variables $m_1$ and $m_2$,
respectively, such that 
\begin{equation} \label{eq:dBianchi-ddKoenigs-U}
\vbn\cdot\vbn \: F^* = U_1 + U_2.
\end{equation}
We will construct one-parameter family of
normal dual congruences conjugated to the lattice $x^*$. The parameter $\lambda$
present 
in the construction will play the role of the spectral parameter of the soliton
theory. 

Consider the following system of equations 
\begin{align} \label{eq:theta-1}
\theta_{(1)} &= \frac{(\theta\vbn +\vbomega)\cdot(\theta\vbn+\vbomega)}
{4\theta(U_1 +\lambda)} ,\\
\label{eq:theta-2}
\theta_{(2)} &= \frac{(\theta\vbn - \vbomega)\cdot(\theta\vbn -\vbomega)}
{4\theta(U_2 -\lambda)} ,
\end{align}
and
\begin{align} \label{eq:lin-w-1}
\Delta_1\bw^* & =\;\;\theta_{(1)}\bx^*_{(1)} -(2 F^*\theta_{(1)}-\theta)\bx^*,\\
\label{eq:lin-w-2}
\Delta_2\bw^* & =   -\theta_{(2)}\bx^*_{(2)} +(2 F^*\theta_{(2)}-\theta)\bx^*,
\end{align}
where $\bw=(\vbomega,\omega^{M+1}):\ZZ^2\to\RR^{M+1}$ and $\theta:\ZZ^2\to\RR$
are unknown fields.
In proving compatibility of the system \eqref{eq:theta-1}-\eqref{eq:lin-w-2}
it is important to notice the following consequences of equations 
\eqref{eq:theta-1}-\eqref{eq:theta-2} and of the constraint
\eqref{eq:dBianchi-ddKoenigs} 
\begin{align}\label{eq:w*n}
\vbomega\cdot\vbn & =  
\theta_{(1)}(U_1 + \lambda) - \theta_{(2)}(U_2-\lambda) ,\\
\label{eq:w*w}
\vbomega\cdot\vbomega & =  2\left( (U_1+\lambda)\theta\theta_{(1)} +
(U_2-\lambda)\theta\theta_{(2)}\right) - \theta^2 \frac{U_1+U_2}{F^*}.
\end{align}
Moreover, in checking compatibility of the system one verifies that $\theta$
satisfies the discrete Moutard equation \eqref{eq:Moutard-dis}.

The dual lattices $y^*_i:\ZZ^2\to\GG_A(M,M+1)$, $i=1,2$, with the
homogeneous coordinates $\by^*_i=(\vbn_i,y_i^{*(M+1)})$ given by
\begin{align} \label{eq:con-y1}
\by^*_{1} &= \; \; \frac{1}{2}(\theta \bx^* - \bw^*) \\
\label{eq:con-y2}
\by^*_{2} &= - \frac{1}{2}(\theta \bx^* + \bw^*),
\end{align}
belong to the pencils containing the dual lines $L^*=x^*\cap w^*$. Equations 
\eqref{eq:lin-w-1}-\eqref{eq:lin-w-2} imply that $\by^*_i$ satisfy the system
\eqref{eq:y1y2}, therefore $y^*_i$ are dual focal lattices of the dual
congruence $L^*$ conjugate to the dual Koenigs lattice $x^*$.

Due to equations \eqref{eq:theta-1}-\eqref{eq:theta-2} the normal vectors 
$\vbn_i$, $i=1,2$, of the dual focal lattices $\by^*_i$, given by equations
\eqref{eq:con-y1}-\eqref{eq:con-y2}, satisfy conditions
\begin{equation*} \label{eq:ni*ni}
\vbn_{2}\cdot \vbn_{2}  = \theta\theta_{(1)}
\left(U_1  + \lambda\right), \quad
\vbn_{1}\cdot \vbn_{1}  = \theta\theta_{(2)} 
\left(U_2  - \lambda\right). 
\end{equation*} 
Then the vectors $\vbn_{1}/\sqrt{\theta\theta_{(2)}}$ and 
$\vbn_{1(1)}/\sqrt{\theta_{(1)}\theta_{(12)}}$
are of equal length, and the hyperplanes
bisecting $y^*_{1}$ and $y^*_{1(1)}$ have the normal vectors
\begin{equation*}
\vb_{1\pm}=\frac{\vbn_{1}}{\sqrt{\theta\theta_{1(1)}}}\pm
\frac{\vbn_{1(1)}}{\sqrt{\theta_{(1)}\theta_{(12)}}}.
\end{equation*} 
Similarly, the hyperplanes
bisecting $y^*_{2}$ and $y^*_{2(2)}$ have the normal vectors
\begin{equation*}
\vb_{2\mp}=\frac{\vbn_{2}}{\sqrt{\theta\theta_{2(2)}}}\pm
\frac{\vbn_{2(2)}}{\sqrt{\theta_{(2)}\theta_{(12)}}}.
\end{equation*}
Equations~\eqref{eq:dBianchi-ddKoenigs-U}, \eqref{eq:w*n}
and \eqref{eq:w*w} imply that
\begin{equation*}
\vb_{1-} \cdot \vb_{2+} = \vb_{1+} \cdot \vb_{2-} =0,
\end{equation*}
which shows that the dual congruence $L^*$ is normal.
\end{proof}
\begin{Cor}
Once the normal dual congruence $L^*$ is found then the homogeneous 
coordinates of the corresponding new dual
Bianchi lattice $x^{*\prime}$ are 
given in terms of $\bw^*$ by 
\begin{equation*} \label{eq:tang-x'}
\bx^{\prime *} = \frac{1}{2} \left[ 
\left( \frac{1}{\theta_{(1)}} - \frac{1}{\theta_{(2)}} \right) \theta\bx^* +
\left( \frac{1}{\theta_{(1)}} + \frac{1}{\theta_{(2)}} \right) \bw^* \right].
\end{equation*} 
\end{Cor}

Following \cite{DNS-I,DNS-gdBs} one can formulate the transition from the normal
vectors $\vbn$ of the dual
Bianchi lattice $x^*$ to the normal vectors $\vbn^\prime$ 
of the new dual Bianchi latice $x^{*\prime}$ as follows.
\begin{Lem} \label{lem:lin-sys-B}
Given $\vbn:\ZZ^2\to\EE^M$ satisfying the discrete Koenigs equation
\eqref{eq:Koenigs-dis-dual} with the constraint \eqref{eq:dBianchi-ddKoenigs-U},
define $\vecbfeta_0=F\vbn$ and supplement it by unit vectors 
$\vecbfeta_A$, $A=1,\dots,M-1$ to an orthogonal basis. Denote by 
$p^B_A$, $q^B_A$,
$A,B=0,1,\dots,M-1$, the functions given by the unique decompositions
\[ \vecbfeta_{A} = \sum_{B=0}^{M-1} p^B_A \vecbfeta_{B(1)}, \quad
\vecbfeta_{A} = \sum_{B=0}^{M-1} q^B_A \vecbfeta_{B(2)},
\]
and let $R = U_1+U_2$, $a=U_1 +\lambda$, 
$b=U_2 -\lambda$.
Then:\\
(i) the linear system
\begin{equation} \label{eq:lin-BE}
\Theta_{(i)}=M_i\Theta, \qquad i=1,2,
\end{equation}
where 
$\Theta=(\theta, \theta^\prime, \theta^{\prime\prime}, y^1,\dots, 
y^{M-1})^T$, and 
\begin{scriptsize}
\begin{align*}
M_1 &= \left( \begin{array}{cccccc}
0&1&0&0&\cdots&0 \\
\frac{R_{(1)}\frac{p_0^0}{F^*}-b}{a_{(1)}}&
\frac{b F^*- R_{(1)}(\frac{R+b}{R}p_0^0-\frac{1}{F^*_{(1)}})}{a_{(1)}}&
\frac{b}{a_{(1)}}(F^*-\frac{R_{(1)}}{R}p_0^0)&
\; \; \frac{R_{(1)}}{a_{(1)}}p_{1}^0& \cdots &
\; \;\frac{R_{(1)}}{a_{(1)}}p_{M-1}^0 \\
-1&F^*&F^*&0&\cdots&0\\
\frac{p_0^{1}}{F^*} & -\frac{R+b}{R}p_0^{1} &
-\frac{b}{R}p_0^{1} & p_{1}^{1} &\cdots & p_{M-1}^{1}\\
\vdots & \vdots & \vdots & \vdots & \ddots & \vdots \\
\frac{p_0^{M-1}}{F^*} & -\frac{R+b}{R}p_0^{M-1} &
-\frac{b}{R}p_0^{M-1} & p_{1}^{M-1} &\cdots & p_{M-1}^{M-1}
\end{array}\right), \\
M_2 &= \left( \begin{array}{cccccc}
0&0&1&0&\cdots &0\\
-1&F^*&F^*&0&\cdots & 0\\
\frac{R_{(2)}\frac{q_0^0}{F^*}-a}{b_{(2)}}&
\frac{a}{b_{(2)}}(F^*-\frac{R_{(2)}}{R}q_0^0)&
\frac{a F^*- R_{(2)}(\frac{R+a}{R}q_0^0-\frac{1}{F^*_{(2)}})}{b_{(2)}}&
-\frac{R_{(2)}}{b_{(2)}}q_{1}^0&\cdots & -\frac{R_{(2)}}{b_{(2)}}q_{M-1}^0
\\
-\frac{q_0^{1}}{F^*}&\frac{a}{R}q_0^{1}&\frac{a+R}{R}q_0^{1}
&q_{1}^{1}&\cdots &q_{M-1}^{1}\\
\vdots & \vdots & \vdots & \vdots & \ddots & \vdots \\
-\frac{q_0^{M-1}}{F^*}&\frac{a}{R}q_0^{M-1}&\frac{a+R}{R}q_0^{M-1}
&q_{1}^{M-1}&\cdots &q_{M-1}^{M-1}
\end{array}\right),
\end{align*}
\end{scriptsize}
is compatible;\\
(ii) the function
\begin{equation*} \label{eq:constraint}
I = (y^1)^2+\dots +(y^{M-1})^2 +\frac{R}{F^*}\theta^2 +
\frac{F^*}{R}\left(
b\theta^{\prime\prime}-a\theta^\prime\right)^2 -
2\theta\left( a\theta^\prime +b\theta^{\prime\prime} \right),
\end{equation*}
is a first integral of the system;\\
(iii) the function $\vbomega=\sum_{A=0}^{M-1} y^A\vecbfeta_A$ with 
\begin{equation}
y^0 = \frac{ a\theta^\prime - b\theta^{\prime\prime}}{R},
\end{equation}
satisfies the linear system
\eqref{eq:lin-w-1}-\eqref{eq:lin-w-2} with $\vbn$ in place of $\bx^*$. 
\end{Lem}
\begin{Cor}
Notice that the solution $\theta$ of the linear system 
\eqref{eq:lin-BE}
is a solution of the discrete Moutard equation \eqref{eq:Moutard-dis}, while
$\theta^\prime=\theta_{(1)}$ and $\theta^{\prime\prime}=\theta_{(2)}$.
\end{Cor}
\begin{Prop}
Given $\vbn:\ZZ^2\to\EE^M$ satisfying the discrete Bianchi 
system \eqref{eq:Koenigs-dis-dual}, 
\eqref{eq:dBianchi-ddKoenigs-U}, let $\theta$ and $\vbomega$
be obtained from the
solution of the linear system \eqref{eq:lin-BE} 
subjected to the admissible constraint $I=0$. Then 
$\vbomega$ satisfies equations \eqref{eq:w*n}-\eqref{eq:w*w} and $\vbn^\prime$,
given by equation~\eqref{eq:tang-x'} with $\vbn$ and $\vbomega$ in place of 
$\bx^*$ and $\bw^*$, is a new solution of the system.
\end{Prop}
\section*{Acknowledgments}
The author would like to thank organizers of the meeting ISLAND~2 for
support during the conference. The paper was supported in part by the
University of Warmia and Mazury grant 522-1307-0201.
\bibliographystyle{amsplain}
\providecommand{\bysame}{\leavevmode\hbox to3em{\hrulefill}\thinspace}
\providecommand{\MR}{\relax\ifhmode\unskip\space\fi MR }
\providecommand{\MRhref}[2]{%
  \href{http://www.ams.org/mathscinet-getitem?mr=#1}{#2}
}
\providecommand{\href}[2]{#2}

\end{document}